\documentclass[a4paper,twocolumn,11pt,accepted=2023-02-22]{quantumarticle}
\pdfoutput=1
\usepackage[utf8]{inputenc}
\usepackage[english]{babel}
\usepackage[T1]{fontenc}
\usepackage{amsmath}
\usepackage{hyperref}

\usepackage{tikz}
\usetikzlibrary{quantikz}
\usepackage{lipsum}
\PassOptionsToPackage{compress}{natbib}
\usepackage[numbers]{natbib}

\usepackage{graphicx}  
\usepackage{subfigure}
\usepackage{multirow}
\usepackage{color}
\usepackage{fancyhdr}
\usepackage{longtable}
\usepackage{parskip}
\usepackage[T1]{fontenc}
\usepackage{dcolumn}   
\usepackage{bm}        
\usepackage{amsfonts}  
\usepackage{amsmath}   
\usepackage{amssymb}   

\begin{document}

\title{Practical computational advantage from the quantum switch on a generalized family of promise problems}

\author{Jorge Escand\'on-Monardes}
\email{jescandon@udec.cl}
\author{Aldo Delgado}
\author{Stephen P. Walborn}

\affiliation{Millennium Institute for Research in Optics and Physics Department, Universidad de Concepci\'on, 160-C Concepci\'on, Chile}


\begin{abstract} 
The quantum switch is a quantum computational primitive that provides computational advantage by applying operations in a superposition of orders. In particular, it can reduce the number of gate queries required for solving promise problems where the goal is to discriminate between a set of properties of a given set of unitary gates. 
In this work, we use Complex Hadamard matrices to introduce more general promise problems, which reduce to the known Fourier and Hadamard promise problems as limiting cases. 
Our generalization loosens the restrictions on the size of the matrices, number of gates and dimension of the quantum systems, providing more parameters to explore. In addition, it leads to the conclusion that a continuous variable system is necessary to implement the most general promise problem. In the finite dimensional case, the family of matrices is restricted to the so-called Butson-Hadamard type, and the complexity of the matrix enters as a constraint. We introduce the ``query per gate'' parameter and  use it to prove that the quantum switch provides computational advantage for both the continuous and discrete cases. 
Our results should inspire implementations of promise problems using the quantum switch where parameters and therefore experimental setups can be chosen much more freely.
\end{abstract}

\maketitle 

\section{Introduction}
Classical events follow well defined causal relations. However, in quantum mechanics one can conceive of a superposition of causal structures, just as quantum systems can be in a superposition of states \cite{Hardy09}. This idea motivated the development of the process matrices formalism \cite{Oreshkov12}, which allows for the description of quantum processes exhibiting an indefinite causal order, i.e., showing correlations incompatible with any definite causal structure.

\begin{figure}[b]
    \centering
    \includegraphics[width=0.4\textwidth]{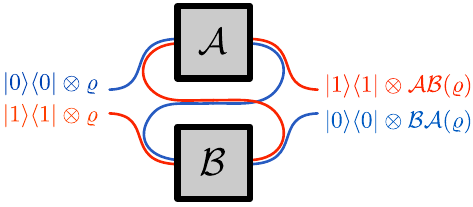}
    \caption{The simplest quantum switch applies two gates in two different orders, which is coherently controlled by a control qubit. In photonic realizations, the order of the operations can be controlled by the path degree of freedom of photons, which we illustrate by colour wires. When the control is in a superposition of states, the order of the gates becomes indefinite.}
    \label{fig:2switch}
\end{figure}
The quantum switch \cite{Chiribella13, Branciard16, Rubino2017, goswami18,Giacomini2016,Bavaresco2019,Cao2022,Wechs21} is a physically realisable instance of indefinite causal order, which consists in the application of two quantum channels $\mathcal{A}$ and $\mathcal{B}$ on a target qubit in two different orders, either $\mathcal{B}\mathcal{A}$ or $\mathcal{A}\mathcal{B}$, coherently controlled by a second qubit (see Fig.\;\ref{fig:2switch}). It has become an interesting resource with applications in quantum computing \cite{Chiribella12,Procopio15,ACB,Taddei,RB22}, quantum communication complexity \cite{Feix15,Guerin16,Wei2019}, communication through noisy channels \cite{Ebler2018,Procopio2019,Procopio2020,Procopio2021,Goswami2020,Guo2020,Chiribella+2021,CWC2021,Wilson2021,Sazim2021}, quantum thermodynamics \cite{Felce2020,Simonov2022} and quantum metrology \cite{Frey2019,Zhao20,CB2021}.

The quantum switch provides computational advantage by reducing the number of queries of unknown quantum channels in a family of tasks, of which the simplest known case is to decide whether a pair of unitary gates commute or anticommute \cite{Chiribella12, Procopio15}. This can be achieved using the quantum switch with a single use of each gate, whilst a fixed-order circuit demands one extra query to one of them. This task has been generalised to the so-called Fourier Promise Problem (FPP). There, a quantum switch of $N$ unitary gates ($N$-\texttt{switch}) deterministically discriminates between a set of $N!$ properties that a set of gates is promised to satisfy upon permutation \cite{ACB}. The $N$-\texttt{switch} coherently applies the $N!$ different orders of the gates such that the permutation property can be determined with a single use of each unitary. On the contrary, the best known fixed-order circuit simulating the quantum switch requires a total of $\mathcal{O}(N^2)$ gate queries \cite{Facchini}. An alternative solution to the FPP using a fixed-order circuit has a query complexity of $\mathcal{O}(N\log N)$, with the cost of using extra ancillary systems \cite{RB21}.

The quantum switch has been experimentally implemented to solve the FPP for $N=2$ qubit gates \cite{Procopio15, Rubino2017}. However, implementation for $N>2$ is technically difficult due to the unfavorable scaling of the dimension of the target system, which must have dimension at least $N!$ \cite{ACB}. While photons do indeed have accessible degrees of freedom with large dimension, such as orbital angular momentum \cite{Mirhosseini_2015} and other types of spatial modes \cite{Neves05,Walborn06}, producing the required transformations even for the smallest dimensional case of $3!$ is challenging. One possible path towards more practical realizations of promise problems was provided in Ref. \cite{Taddei}, where the Hadamard Promise Problem (HPP) was introduced, with an explicit solution  experimentally demonstrated using a quantum switch that applied four sequences of four gates to a qubit target system. A more general definition of HPPs using $N$ gates was recently introduced in Ref. \cite{RB22}, with the same distinctive feature of requiring just a qubit as the target system. Both the FPP and HPP can be cast as the task of determining which column of a matrix describes the permutation properties of a set of unitaries, relating the promise satisfied by the set of unitaries to the entries of a Fourier or Hadamard matrix.  

In this paper, we introduce a generalized set of promise problems, which in addition to providing much more freedom in regards to the dimension of the target system, uncouples the order of the matrix and the number of quantum gates, offering more parameters to explore. To achieve this, we turn to the more general Complex Hadamard (CH) matrices to define the Complex Hadamard Promise Problem (CHPP), providing a large family of promise problems. The CHPP reduces to the FPP or HPP as limiting cases. In regard to discrete variable target systems, our approach removes the unfavorable factorial scaling of the FPP, and we find that CHPPs exist for every finite dimension with some additional constraints on the entries of the CH matrix. In addition, the entire family of CHPPs can be defined for continuous variable (CV) target systems. Moreover, some CHPPs are exclusively defined for CV systems, which could possibly lead to new applications. For a comparison with related works on these features, see Table I. Looking towards implementation, we show how the quantum switch solves the CHPP and offers computational advantage compared to the best known fixed-order circuits for both discrete and continuous variable target systems. This result renders new flexibility to the design of experiments using the quantum switch to solve promise problems and opens the possibility of using CV platforms to this end.

\begin{table*}[t]
    \centering
    {\footnotesize
    \begin{tabular}{|c|c|c|c|c|c|}
    \hline
&&Araújo,&Taddei&Renner,&\\
Parameter&Chiribella\cite{Chiribella12}&Costa,&et al.\cite{Taddei}&Brukner\cite{RB22}&This work\\
&&Brukner\cite{ACB}&&&\\
\hline
Matrix $M$ & $H_2$ & $F$ & $H_4$ & $H$ & $CH$\\

Number $N$ of gates & $2$ & $N\geq 2$ & $4$ & $N\geq 2$ & $N\geq 2$\\

Dimension $d$ of the target system & $2$ & multiple of $N!$ & $2$ & $2$ & $d\geq 2$ and CV\\

Number $p$ of permutations & $2$ & $N!$ & $4$ & $2^{N-1}$ & $p\leq N!$\\
\hline
Subclass & HPP, & FPP & HPP & HPP & BHPP,\\
& FPP &&&& CHPP\\
\hline
    \end{tabular}
    }
    \label{Summarytable}
    \caption{\textbf{Promise problems in the literature:} Comparison between different promise problems and explicit solutions using the quantum switch. The last row shows the subclass of promise problems covered in each related work.}
\end{table*}

The structure of the article is as follows. Section \ref{sec CH matrices} introduces Complex Hadamard matrices and some relevant properties.  The CHPP is defined in Section \ref{secCHPP} and shown to be well defined for every dimension, with some restrictions on the CH matrices. Section \ref{secSolQS} shows the solution of the CHPP using the quantum switch and Section \ref{secFixSol} assesses the computational advantage of the quantum switch against the best known fixed-order solutions. A discussion on possible experimental implementations is found in Section \ref{secDiscuss}, and concluding remarks are drawn in Section \ref{secConcl}.

\section{Complex Hadamard matrices}
\label{sec CH matrices}

A square matrix $M$ of size $p$ is called a \textit{Complex Hadamard} (CH) matrix if its entries are unimodular ($|M_{jk}|=1$) and $MM^\dagger=pI$, where $I$ is the identity matrix of size $p$ (for further details see \cite{Tadej06}). We denote the set of complex Hadamard matrices of size $p$ as $CH(p)$. It is clear from the definition that if $M\in CH(p)$, then $M/\sqrt{p}$ is unitary, and that any two columns or rows of $M$ are orthogonal. If the entries of a complex Hadamard matrix in the first column and the first row are all equal to one, we say that the matrix is in its \textit{dephased} form. Any $M\in CH(p)$ can be cast in dephased form\footnote{For any Complex Hadamard matrix $M$, there exist diagonal unitary matrices $D_1$ and $D_2$ such that $D_1 M D_2$ is dephased.}.

Since the entries of a complex Hadamard matrix are unimodular, they can be written as
\begin{equation}
M_{jk}=e^{i\phi_{jk}}\;.
\end{equation}
The phases $\phi_{jk}\in [0,2\pi)$ are the entries of a matrix $\phi$ known as a \textit{log-Hadamard} matrix. For example, if $M$ is dephased, then all the entries of the first row and first column of the corresponding log-Hadamard matrix are equal to zero.

A special subset of complex Hadamard matrices is the set of \textit{Butson-type complex Hadamard matrices}, or Butson-Hadamard (BH) matrices for short \cite{Butson}. They satisfy the additional condition that all their entries are roots of unity. We say that $M\in CH(p)$ is a BH matrix of \textit{complexity} $d$ if all its entries are $d$-th roots of unity, what we denote as $M\in BH(p,d)$. Notice that the log-Hadamard elements associated to a BH matrix of complexity $d$ have the form $\phi_{jk}=2\pi q_{jk}/d$, with $q_{jk}\in\{0,1,...,d-1\}$.

Particular examples of BH matrices are Fourier matrices $F_d$, belonging to $BH(d,d)$ with entries
\begin{equation}
(F_d)_{jk}=e^{i\frac{2\pi jk}{d}}\;,
\end{equation}
where we have dropped the normalizing constant. Another known family of BH matrices is the set of \textit{real Hadamard} matrices, with entries equal to $\pm 1$, hence corresponding exactly to the set $BH(p,2)$.

It is important to notice that not every CH matrix is BH. As an example, let us consider the case of CH matrices of size $4$. Every matrix in $CH(4)$ is equivalent \footnote{Two complex Hadamard matrices $M_1$ and $M_2$ are said to be equivalent if there are permutation matrices $P_1$, $P_2$ and diagonal matrices $D_1$, $D_2$ such that $M_1=D_1 P_1 M_2 P_2 D_2$.} to a matrix in the family
\begin{equation}
    F_4^{(1)}(a)=\begin{pmatrix}
1 & 1 & 1 & 1\\
1 & ie^{ia} & -1 & -ie^{ia}\\
1 & -1 & 1 & -1\\
1 & -ie^{ia} & -1 & ie^{ia}
\end{pmatrix},
\label{CH4matrix}
\end{equation}
with $a\in[0,\pi)$. Conversely, every matrix in this family is a CH. Notice that $F_4^{(1)}(0)$ is the Fourier matrix of size $4$, $F_4$, while $F_4^{(1)}(\pi/2)$ is a real Hadamard matrix. To get a CH matrix which is not BH, it is enough to write $a=2\pi\nu$ and fix $\nu$ as an irrational number.

\section{Complex Hadamard Promise Problem}\label{secCHPP}
\subsection{The Problem}
In this section we introduce the Complex Hadamard Promise Problem (CHPP). Our definitions follow those laid out in Ref. \cite{ACB}.

Let $M$ be a $CH(p)$ matrix in its dephased form and consider a set of $N$ unknown unitary gates $U_0,...,U_{N-1}$. We define the product $\Pi_0=U_{N-1}U_{N-2}...U_0$ and denote different permutations of the same gates by $\Pi_1,...,\Pi_{p-1}$. The Complex Hadamard Promise states that the set of unitaries satisfies the following property for one of the columns of $M$:
\begin{equation}
    \forall j\in \{0,...p-1\}: \Pi_j=M_{jk}\cdot \Pi_0.
    \label{eq:CHPromise}
\end{equation}
The problem is to find the column $k$.

Notice that the size $p$ of the matrix defines the number of permutations that will be considered. This sets a lower bound for the number of unitaries $N$, which must be such that:
\begin{equation}
    p\leq N!.
\end{equation}

\subsection{Existence of unitaries satisfying the promise: Finite dimensional target}

The definition of the CHPP does not garantee the existence of a set of unitaries satisfying the promise under some permutations for a given CH matrix. Here, we prove that those unitaries do exist, but some restrictions appear on the CH matrices if we choose unitary gates acting on finite dimensional systems.

Let us consider a $D$-dimensional target system. In that case, gates are $D\times D$ unitary matrices. We can follow a similar approach as in Ref. \cite{ACB} to find that $M_{jk}^D=1$, due to $\mathrm{det} \Pi_j=M_{jk}^D\;\mathrm{det} \Pi_0$ and $\mathrm{det} \Pi_j=\mathrm{det} \Pi_0$,  for any $j$ and $k$. Consequently, $M$ must be a BH matrix. More  specifically, a CHPP can be implemented in finite dimensional systems only if the CH matrix specifying the promise is Butson-type with complexity $d=D/m$ for some positive integer $m$. Moreover, if a CHPP can be formulated for a $d$-dimensional target, it can also be formulated for any system with dimension that is a multiple of $d$. This follows from the definition of BH matrices, which implies that $BH(p,d)\subseteq BH(p,md)$. For the canonical case of a qubit target, the CHPP must be specified by a real Hadamard matrix, i.e., the CHPP gets restricted to the HPP already introduced in Refs. \cite{Taddei,RB22}, which may also be implemented in any even dimensional target system.

It is useful to set some nomenclature at this point. We will refer to particular classes of CHPPs, such as Butson-Hadamard Promise Problems (BHPPs), Fourier Promise Problems (FPPs) and Hadamard Promise Problems (HPPs), when the CH matrices specifying them are Butson-type, Fourier and Hadamard, respectively. Notice that the original definition of the FPP \cite{ACB} uses all possible permutations of $N$ unitaries, hence the dimension of the target must be a multiple of $d=N!$. Since we are loosening the number of permutations required for the promise, our definition allows the FPP to be formulated with Fourier matrices whose size, and therefore the dimension of the target system, does not need to be as large as the factorial of the number of unitaries, allowing for more realistic experimental scenarios. Similarly, Ref. \cite{RB22} introduced a method to construct HPPs from simpler HPPs. It leads to specific relations between $p$ and $N$ depending on the fundamental HPP considered. Their explicit construction is based on Chiribella's task \cite{Chiribella12} and leads to $p=2^{N-1}$, which is included in Table I. It can be proven that Taddei et al.'s task \cite{Taddei} is also fundamental in this sense, providing a different scaling for $p$ in terms of $N$. With our approach instead, we shall be able to build solutions for every HPP, be it fundamental or not.

Now, let us show that any BHPP is well defined on a target with compatible dimension, i.e. we can always find sets of unitary gates fulfilling the promise when the dimension of the target is multiple of the complexity of the BH matrix. The construction is as follows:

Let us consider the generalized Pauli $X$ and $Z$ gates, defined as:
\begin{align}
    X&:=\sum_{j=0}^{d-1}\ket{j\oplus 1}\bra{j},\\
    Z&:=\sum_{j=0}^{d-1}\omega_d^j\ket{j}\bra{j},
\end{align}
where $\oplus$ denotes sum modulo $d$ and $\omega_d=\exp(2\pi i/d)$ is the primitive $d$-th root of unity.
These unitary operators satisfy $X^d=Z^d=I_d$ and the commutation relation 
\begin{equation}
    ZX=\omega_d XZ\;,
\end{equation}
which can be extended using mathematical induction to
\begin{equation}
    Z^j X^k=\omega_d^{jk}X^k Z^j\;.
    \label{ZXcom}
\end{equation}
Now, let us define the following $p$ unitaries:
\begin{equation}
    \begin{array}{rclc}
    U_0&=&X\;, & \\
    U_j&=&Z^{q_j}\;,&\forall j=1,...,p-1\;,
    \end{array}
\end{equation}
where $q_j$ are integers, and consider the following $p$ permutations, where $U_0$ is shifted one position to the left each time:
\begin{equation}
    \begin{array}{rcl}
    \Pi_0&=&U_{p-1}U_{p-2}\cdots U_1U_0\;,\\
\Pi_1&=&U_{p-1}U_{p-2}\cdots U_0U_1\;,\\
 \vdots &&\vdots \\
\Pi_{p-2}&=&U_{p-1}U_{0}\cdots U_2U_1\;,\\
\Pi_{p-1}&=&U_{0}U_{p-1}\cdots U_2U_1\;.
    \end{array}
    \label{permut}
\end{equation}
Using the commutation relation \eqref{ZXcom} it is easy to show that
\begin{equation}
    \Pi_j=e^{i\frac{2\pi}{d}(q_1+q_2+...+q_j)}\Pi_0\;,\;\forall j=1,...,p-1\;.
\end{equation}
Hence, given $M\in BH(p,d)$ dephased and an arbitrary column
\begin{equation}
M_{*,k}=(1,e^{i\frac{2\pi}{d}q_{1k}}, e^{i\frac{2\pi}{d}q_{2k}}, \cdots,  e^{i\frac{2\pi}{d}q_{(p-1)k}})^T
\end{equation}
of $M$, we just need to choose $q_j=q_{jk}-q_{(j-1)k}$ for $j=1,...,p-1$ to obtain a set of $p$ unitaries satisfying the promise, as we wanted to show.

\subsection{Existence of unitaries satisfying the promise: Continuous variable target}\label{CVunitaries}

As mentioned in Section \ref{sec CH matrices}, there are Complex Hadamard matrices which are not Butson-type. Consequently, they do not define valid CHPPs on finite dimensional target systems. Interestingly, the restriction imposed by the determinant in the finite dimensional case is dropped in the continuous variable (CV) regime. Moreover, as we next prove, unitary gates satisfying the promise do exist for every CH matrix in its dephased form and for each of its columns, thus the CHPP is always well defined for CV target systems.

The construction of unitaries is analogous to that in previous section. Now let us consider the displacement operators
\begin{equation}
\begin{array}{rclr}
     X_{\alpha}&=&e^{-i\alpha \hat{p}}&,  \\
     Z_{\beta\gamma}&=&e^{i(\beta\hat{x}+\gamma \hat{p})}&, 
\end{array}
\end{equation}
where $\alpha, \beta, \gamma$ are real numbers, while $\hat{x}$ and $\hat{p}$ are position and momentum operators satisfying $[\hat{x},\hat{p}]=iI$. Using the Baker-Campbell-Hausdorff formula we have
\begin{equation}
    Z_{\beta\gamma} X_\alpha=e^{i\alpha\beta}X_\alpha Z_{\beta\gamma}\; .
\end{equation}
Setting the gates
\begin{equation}
    \begin{array}{rclc}
    U_0&=&X_\alpha \;,& \\
    U_j&=&Z_{\beta_j\gamma_j}\;,&\forall j=1,...,p-1\; ,
    \end{array}
\end{equation}
and considering the same $p$ permutations of \eqref{permut}, it can be shown that
\begin{equation}
    \Pi_j=e^{i\alpha(\beta_1+\beta_2+...+\beta_j)}\Pi_0\; ,\; \forall j=1,...,p-1\;.
\end{equation}
Given $M$ dephased and an arbitrary column
\begin{equation}
M_{*,k}=(1, e^{i\phi_{1k}}, e^{i\phi_{2k}}, \cdots,  e^{i\phi_{(p-1)k}})^T
\end{equation}
of $M$, we can choose $\alpha\neq 0$, $\beta_j=(\phi_{jk}-\phi_{(j-1)k})/\alpha$ and $\gamma_j$ arbitrary for $j=1,...,p-1$ to have a set of unitaries satisfying the promise.

Notice that if $\{U_j\}$ is a set of unitaries satisfying a specific promise and $V$ is any unitary gate, then the set $\{VU_jV^\dagger\}$ fulfills the same promise.

\subsection{Existence of unitaries satisfying the promise: Minimal sets}\label{minimalsets}

From the definition of the problem, we already know that the number  of unitaries $N$ required to satisfy a promise specified by a $CH(p)$ matrix is such that $p\leq N!$. We will say that a set of unitaries satisfying the promise is \textit{minimal} if the number of unitaries is the minimum $N$ such that $p\leq N!$. In this section we discuss the following question: is it possible to define a minimal set of unitaries satisfying a given promise?

Some examples are already known. Indeed, Ref. \cite{ACB} shows minimal sets of unitaries for some FPPs; since they consider only matrices of order $N!$, their construction of unitaries is trivially minimal. Also, in \cite{RB22} a specific HPP is considered for any $N$, with a construction of unitaries that is minimal for $N=2$, $3$ and $4$, but not for larger $N$. Meanwhile, our construction above is minimal only for matrices of size $p=2$ and $3$.

We also investigated the case of a CHPP specified by a $CH(4)$ matrix. In particular, we considered an arbitrary member of the parameterised family defined in  Eq. \eqref{CH4matrix}. Since this family includes matrices which are not Butson-type, we required unitary gates acting on a CV system. This promise consists on four permutations; therefore, a minimal set of unitaries must have three gates: $U_0$, $U_1$ and $U_2$. We chose $\Pi_0=U_2U_1U_0$, $\Pi_1=U_2U_0U_1$, $\Pi_2=U_0U_2U_1$ and $\Pi_3=U_0U_1U_2$, and the ansatz $U_0=X_{\alpha_0}Z_{\beta_0}$, $U_1=X_{\alpha_1}Z_{\beta_1}$ and $U_2=X_{\alpha_2}Z_{\beta_2}$. Solving for $\alpha_j$ and $\beta_j$, we found solutions for each column and any $a\in[0,\pi)$. Some solutions are the following:
\begin{itemize}
    \item \textbf{For column $k=0$:} $\alpha_j=0$ and $\beta_j$ arbitrary for every $j$.
    \item \textbf{For column $k=1$:} $\alpha_0=0$, $\alpha_1\neq 0$ arbitrary, $\alpha_2=(\pi-2a)\alpha_1/(\pi+2a)$, $\beta_0=(\pi+2a)/2\alpha_1$, $\beta_1$ arbitrary and $\beta_2=(3\pi+2\alpha_2\beta_1-2a)/2\alpha_1$.
    \item \textbf{For column $k=2$:} $\alpha_0=0$, $\alpha_1\neq 0$ arbitrary, $\alpha_2=-\alpha_1$, $\beta_0=\pi/\alpha_1$, $\beta_1$ arbitrary, and $\beta_2=-\beta_0-\beta_1$.
    \item \textbf{For column $k=3$:} If $a=\pi/2$, then $\alpha_0\neq 0$ arbitrary, $\alpha_1=0$, $\alpha_2$ arbitrary, $\beta_0$ arbitrary, $\beta_1=0$, $\beta_2=(-\pi+\alpha_2\beta_0)/\alpha_0$.
    If $a\neq \pi/2$, then $\alpha_0=0$, $\alpha_1\neq 0$ arbitrary, $\alpha_2=(-3\pi+2a)\alpha_1/(\pi-2a)$, $\beta_0=(-\pi+2a)/2\alpha_1$, $\beta_1$ arbitrary, and $\beta_2=(\pi+2\alpha_2\beta_1-2a)/2\alpha_1$.
\end{itemize}
Thus, for every CHPP of size 4, minimal sets of unitaries satisfying the promise do exist.

We conjecture that minimal sets of unitaries satisfying a CHPP exist for every size $p$. However, they may require joint systems as target, just like the construction in Ref. \cite{ACB}. That is, for some $p$, keeping the number of unitaries low would demand an increase in the dimension of the target.

\section{Solution using the quantum switch}\label{secSolQS}

\begin{figure}[bt]
    \centering
    \includegraphics[width=0.48\textwidth]{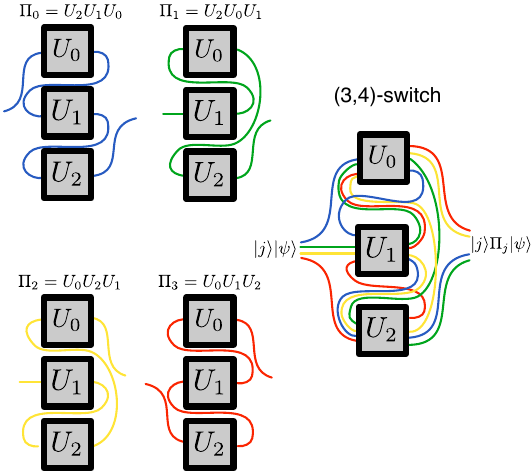}
    \caption{An $(N,p)$-\texttt{switch} applies $N$ unitaries in $p$ different orders. In the image we show a $(3,4)$-\texttt{switch}, where the colour wires represent the states of the control system.}
    \label{fig:34switch}
\end{figure}

The quantum switch $S$ is a device that applies a set of unknown gates on a target quantum system in different orders that are determined by a control quantum system. For each basis state $\ket{j}$ of the control, a corresponding permutation $\Pi_j$ of the gates is applied on the target, which is prepared in some arbitrary initial state $\ket{\psi}$. Mathematically,
\begin{equation}
    S\left(\ket{j}\otimes\ket{\psi}\right)=\ket{j}\otimes\Pi_j\ket{\psi}\;.
\label{qswitch}
\end{equation}
If the control is in a superposition of states, then the gates are applied in a superposition of orders (see Fig. \ref{fig:34switch}).

In the literature, it is common to find the expression "$N$-\texttt{switch}" for a quantum switch of $N$ gates. As we mentioned above, we do not necessarily require all possible permutations, so it makes sense to refer to an $(N,p)$-\texttt{switch}, i.e., a quantum switch of $N$ gates applied in $p$ different orders. Of course, since there are different combinations of $p$ permutations of $N$ elements, the expression "$(N,p)$-\texttt{switch}" properly denotes a \textit{class} of quantum switches. Therefore, the $N$-\texttt{switch} considering all the $N!$ permutations, as originally introduced in Ref. \cite{ACB}, is the only quantum switch in the class of $(N,N!)$-\texttt{switches}. Rigorously speaking, the HPP implemented in Ref. \cite{Taddei} is solved using a $(4,4)$-\texttt{switch} and the scaling method proposed in Ref. \cite{RB22} uses $(N,2^{N-1})$-\texttt{switches} that can be described as nested $(2,2)$-\texttt{switches}. Note that the exploration of different combinations of permutations of channels was also highlighted in Refs. \cite{Procopio2020,Wilson2021,Sazim2021}.

The quantum switch plays a key role in the quantum approach to  solving the CHPP, which is a straightforward extension of the protocol for solving the FPP in Ref. \cite{ACB}. It is as follows: First, consider a CHPP specified by a matrix $M\in CH(p)$ and a set of $p$ permutations $\{\Pi_j\}$ of $N$ unitaries. Suppose the unitaries satisfy the promise for the $k$-th column of $M$. Now initialize a control system of dimension $p$ in the state $\ket{0}$, and a target system of dimension compatible with the problem in an arbitrary state $\ket{\psi}$.
The unitary gate $M/\sqrt{p}$ is applied on the control system. Since $M$ must be in its dephased form, the joint state becomes
\begin{equation}
\left(\frac{1}{\sqrt{p}}M\otimes I\right)\left(\ket{0}\otimes\ket{\psi}\right)=\frac{1}{\sqrt{p}}\sum_{j=0}^{p-1}\ket{j}\otimes\ket{\psi} \;.
\end{equation}
Then, applying the corresponding quantum switch we get
\begin{align}
S\left(\frac{1}{\sqrt{p}}\sum_{j=0}^{p-1}\ket{j}\otimes\ket{\psi}\right)&=\frac{1}{\sqrt{p}}\sum_{j=0}^{p-1}S\left(\ket{j}\otimes\ket{\psi}\right)\nonumber\\
&=\frac{1}{\sqrt{p}}\sum_{j=0}^{p-1}\ket{j}\otimes \Pi_j\ket{\psi}\nonumber\\
\end{align}
where we used Eq. \eqref{qswitch}. Using now the definition of the promise \eqref{eq:CHPromise} and recognizing the matrix multiplication, we have 
\begin{align}
S\left(\frac{1}{\sqrt{p}}\sum_{j=0}^{p-1}\ket{j}\otimes\ket{\psi}\right)&=\frac{1}{\sqrt{p}}\sum_{j=0}^{p-1}M_{jk}\ket{j}\otimes\Pi_0\ket{\psi}\nonumber\\
&=\frac{1}{\sqrt{p}}M\ket{k}\otimes\Pi_0\ket{\psi}.
\end{align}
Applying the inverse $M^\dagger /\sqrt{p}$ on the control,  the number $k$ of the column becomes encoded on the control system:
\begin{equation}
\left(\frac{1}{\sqrt{p}}M^\dagger\otimes I\right) \frac{1}{\sqrt{p}}M\ket{k}\otimes\Pi_0\ket{\psi}
=\ket{k}\otimes\Pi_0\ket{\psi}.
\end{equation}
Finally, a projective measurement on the control system gives us the solution of the problem deterministically.

The solution of the CHPP using the quantum switch only requires a single use of each gate. Indeed, this can be shown by coupling one high-dimensional ancilla state $\ket{f}$ as "counter" to each gate \cite{ACB}, whose state evolves from $\ket{f}$ to $\ket{f+1}$ each time the corresponding gate is used, with $f\in\mathbb{N}$ . More specifically, if the counters are initialized in state $\ket{0}$, then the state of each counter becomes $|1\rangle$ after applying the quantum switch, since the unitaries are coherently controlled. The solution using the quantum switch is the most efficient solution known to the CHPP, as we discuss in next section.

\section{Solution using fixed-order circuits}\label{secFixSol}
\begin{figure*}[t]
    \centering
    \begin{quantikz}[row sep={0.7cm,between origins}, column sep={0.4cm}]
        \lstick[wires=2]{$|j\rangle$}&&\lstick{$|c_0\rangle$} & \ctrl{3} & \qw & \ctrl{3} & \qw & \qw & \qw & \qw & \octrl{3} & \qw & \octrl{3} & \qw & \qw & \qw & \qw\rstick[wires=3]{$\sum_j\ket{j}\Pi_j\ket{\psi}$}
        \\
        &&\lstick{$|c_1\rangle$} & \qw & \qw & \qw & \ctrl{3} & \qw & \ctrl{3} & \qw & \qw & \qw & \qw & \octrl{3} & \qw & \octrl{3} & \qw
        \\
        &&\lstick{$|\psi\rangle$} & \targX{} & \gate{U_0} & \targX{} & \targX{} & \gate{U_1} & \targX{} & \gate{U_2} & \targX{} & \gate{U_0} & \targX{} & \targX{} & \gate{U_1} & \targX{} & \qw
        \\
        &&\lstick{$|a_0\rangle$} & \swap{-1} & \qw & \swap{-1} & \qw & \qw & \qw & \qw & \swap{-1} & \qw & \swap{-1} & \qw & \qw & \qw & \qw \rstick{$U_0\ket{a_0}$}
        \\
        &&\lstick{$|a_1\rangle$} & \qw & \qw & \qw & \swap{-2} & \qw & \swap{-2} & \qw & \qw & \qw & \qw & \swap{-2} & \qw & \swap{-2} & \qw \rstick{$U_1\ket{a_1}$}
    \end{quantikz}
    \caption{Here we show a fixed-order simulation of a $(3,4)$-switch. The arrangement of the $U_i$ gates in the circuit is the shortest one containing the permutations 012, 021, 120 and 201. This circuit has a two-qubit control register, one target and two ancillary systems. The state $|j\rangle$ of the control defines which permutation will be applied to the target system $|\psi\rangle$. The target is routed with the aid of controlled-swap gates which interchange the target and one of the ancillas $|a_i\rangle$.}
    \label{fig:Circuitscs}
\end{figure*}
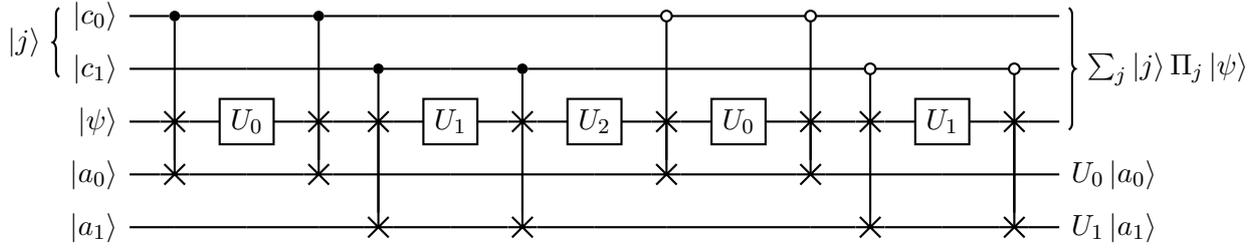
The CHPP can also be solved using fixed-order circuits. However, as we discuss next, the solution using the quantum switch requires a fewer number of gate queries than the best known fixed-order alternatives, which is referred to as \textit{query complexity advantage}.

\subsection{Simulation of the quantum switch}

A first way to solve the CHPP using a fixed-order circuit is by simulating the quantum switch in the circuit model \cite{Facchini,Colnaghi12}. The best simulation known requires as many gates as elements are in the shortest common supersequence (SCS)\footnote{Given a set of sequences $s_1,...,s_r$, a supersequence $s'$ is a sequence  such that every $s_i$ can be recovered by deleting some elements from $s'$. A SCS is a supersequence of minimal length and may not be unique. For example, $s'=0120210$ is a SCS for the set of sequences $\{012, 021, 102, 120, 201, 210\}$.} of the permutations involved \cite{Facchini}. The circuit consists of a target and a $p$-dimensional control system. The gates are applied in the order specified by the SCS and for each extra use of a unitary gate an ancillary system with the same dimension of the target is added \cite{Taddei}. Controlled-swap gates route the target in such a way that it undergoes the required permutation of the gates. Furthermore, the ancillas are used in such a way that they end up being disentangled from the rest of the system. See Fig.\;\ref{fig:Circuitscs} for a $(3,4)$-\texttt{switch} fixed-order simulation example.

\begin{figure*}[t]
    \centering
    \includegraphics[width=\linewidth]{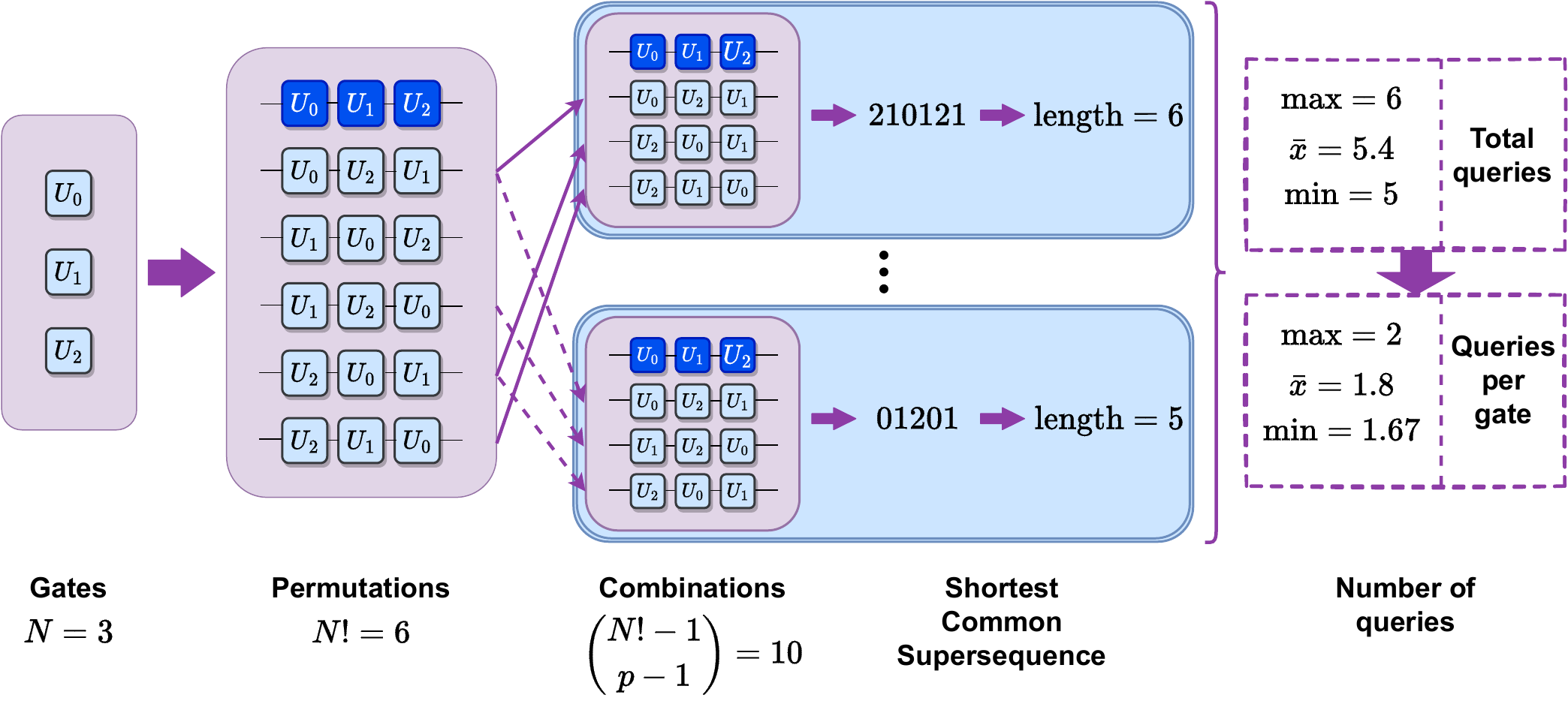}
    \caption{We illustrate how to calculate numerically the number of queries required by the simulation of a $(3,4)$-\texttt{switch}. Firstly, all the permutations of the gates and all the combinations of $p$ permutations are computed. We demand the combinations to include $\Pi_0=U_{N-1}...U_1U_0$, otherwise the gates are relabeled. Then, one SCS for each combination is found numerically and their lengths are averaged.}
    \label{fig:Combinations}
\end{figure*}

In order to assess the query advantage of the quantum switch against its fixed-order simulation, we have to find the length of the corresponding SCS, which will be equal to the total number of queries. However, the problem of finding a SCS has been proved to be NP-complete \cite{Raiha} and some approximate algorithms have been reported in the literature (for further details see Ref. \cite{Ning}). In the case of the SCS containing all the $N!$ permutations of $N$ elements, its length grows with $\mathcal{O}(N^2)$ \cite{Facchini} and an upper bound is set by the length of the Koutas-Hu supersequence \cite{Koutas}. In this work, we deal with a broader scenario considering $p\leq N!$ permutations, hence the SCS may be shorter. To find it, we proceed by numerical exhaustive search, i.e., we list and sort by length all the sequences shorter than the Koutas-Hu supersequence and for each one we decide if it contains or not the desired set of permutations; the search halts when the first supersequence is found and its length is recorded. In this fashion, we explored the cases $N=3$ and $N=4$ with varying $p$, confirming that the total number of queries depends on both $N$ and $p$, and also on the specific set of permutations. For example, simulating a $(3,4)$-\texttt{switch} demands 5 or 6 gate uses, which depends on the set of chosen permutations, with an average of 5.4 queries (see Fig. \ref{fig:Combinations}). For the case of the $(3,3!)$-\texttt{switch}, 7 calls to the gates are required. Similarly, simulating a $(4,4)$-\texttt{switch} requires a number of gate uses in a range from 6 to 9, with an average of 7.43, while simulating all the permutations of 4 gates demands 12 queries. 

\begin{figure}[t]
    \centering
    \includegraphics[width=\linewidth]{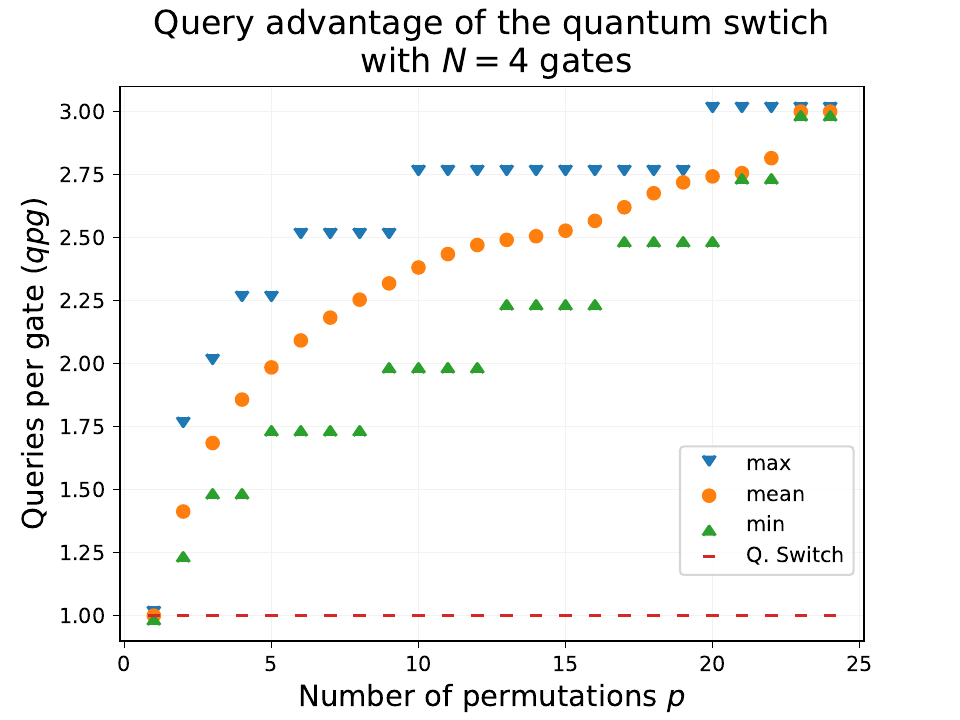}
    \caption{The graph shows the minimum, maximum and average queries per gate ($qpg$) of a fixed-order simulation of an $(N,p)$-\texttt{switch}, for $N=4$ and varying $p$. The dashed line shows the performance of the quantum switch.}
    \label{fig:lengthsSCS}
\end{figure}

In Refs. \cite{ACB,RB22}, each promise problem is specified by the number of gates $N$. Here, we have seen that the same CHPP can be defined for different values of $N$. As a consequence, the total number of queries may not be the best parameter to assess the computational advantage of the quantum switch over solutions based on fixed-order circuits solving the same task. We find it useful to introduce the average number of \textit{queries per gate} ($qpg$) as a parameter for query complexity advantage. The quantum switch has a constant value of $qpg_{switch}=1$ for every CHPP, while its simulation via the shortest common supersequence grows with $qpg=\mathcal{O}(N)$. The simulations of the $(3,4)$- and $(4,4)$-\texttt{switches} described in the previous paragraph have averages $qpg_{(3,4)}=1.8$ and $qpg_{(4,4)}=1.86$, respectively, and if we consider all the permutations of 3 and 4 gates, the corresponding simulations of the quantum switch will be $qpg_{(3,3!)}=2.33$ and $qpg_{(4,4!)}=3$. A thorough description for the case $N=4$ is shown in Fig. \ref{fig:lengthsSCS}. It can be seen that the quantum switch offers advantage for every $p\geq 2$ with the greatest difference for $p=N!$.

\subsection{Other approaches} 

We could ask whether different approaches using fixed-order circuits could solve any arbitrary CHPP as well. In particular, tomographic reconstruction of the gates is discussed in Ref. \cite{Taddei}, with cubically worse performance in $N$ than with the quantum switch; similarly, direct reconstruction of the permutations is still quadratically worse than the quantum switch. Alternatively, the authors of Ref. \cite{RB21} proposed a causal circuit which solves the original FPP with $\mathcal{O}(N\log N)$ queries to the gates ($qpg=\mathcal{O}(\log N)$), and in Ref. \cite{RB22} they extended it to solve the HPP. That is the best fixed-order circuit known to solve these particular tasks for large $N$ and it is conjectured that no causal algorithm can solve them more efficiently. It is an open question whether this approach can be extended to the general CHPP.

There is one additional scenario that deserves special attention, since it could solve the CHPP also with $qpg=1$. It is known from the $d$-dimensional dense coding protocol that we can send a 2-dit message using 1 qudit system when a maximally entangled high-dimensional channel has been previously shared. This information is encoded by applying a generalized Pauli gate on the system and retrieved at the end in a Bell-state measurement \cite{Liu}. Retrieving the message is equivalent to discriminating which Pauli gate was applied. Consequently, to solve a CHPP where the unitary gates are known to be a certain set of Pauli gates, we could identify each unitary by using it to perform a dense coding protocol and then calculate their permutations; it would solve the CHPP with a single query to each gate, the same as the quantum switch. Notwithstanding, this approach consumes $N$ pairs of maximally entangled qudits (one for each unitary) and requires prior information about the gates. Also, the Bell-state measurement must be performed with respect to the correct basis. Instead, the quantum switch based solution does not consume entanglement nor need prior information about the gates.

\section{Discussion}\label{secDiscuss}
The introduction of the more general CHPP opens the possibility for experimental realizations of the computational advantage of the quantum switch in higher dimensions, both in terms of the number of permutations and the dimension of the target system. In addition, the more general description via CH matrices should allow these problems to be more readily adapted to different experimental platforms that do not necessarily implement a Fourier or Hadamard transformation directly.   

All experimental implementations of the quantum switch have been realized in photonic systems, and most of them have exploited the polarization degree of freedom as either the control or target system \cite{Procopio15,Rubino2017,goswami18,Wei2019,Guo2020,Taddei}. For higher-dimensional target systems, implementations of the BHPP with the quantum switch requires two photonic degrees of freedom (DOF) with dimension greater than two. Due to the fact that arbitrary $d\times d$ unitary operations can be constructed from $2 \times2$ beam splitters and phase shifters \cite{Reck94}, the path DOF is a natural candidate for the control system. Moreover, multiport beam splitters can be used to implement Hadamard and/or Fourier transformations directly \cite{crespi16,carine20}.
For the target system, a number of transverse DOFs could be used.  For example, Heisenberg-Weyl operators can be implemented on the orbital angular momentum of photons using spiral wave plates and other linear devices \cite{Palici2020}. $D$-dimensional states and unitaries can also be implemented using gratings and near-field diffraction via the Talbot effect \cite{farias15}, an approach which could also apply to matter waves \cite{barros17}. Regarding the implementation of the most general CHPP, a CV target system is required. Particularly, simple photonic realizations could exploit the continuous transverse position/momentum of photons, where the appropriate Heisenberg-Weyl operators can be implemented using lenses and phase devices \cite{tasca11}. 

Our results might also inspire exploration of promise problems and the quantum switch using non-photonic systems. As noted in Ref. \cite{Zhao20}, there are a number of systems with continuous degrees of freedom that could be candidates for the quantum switch, including trapped ions or cavity QED systems. 

Generalization of the Fourier and Hadamard promise problem to the CHPP also inspires further study from a fundamental point of view. Causal non-separability of the quantum switch has been experimentally confirmed for the case of two gates acting on a qubit target \cite{Rubino2017, goswami18,Cao2022}, but a similar proof with gates acting on either higher-dimensional or CV systems is still lacking. Furthermore, as far as we are aware, nor has experimental demonstration of causal non-separability of the $(N,p)$-\texttt{switch} been reported.

\section{Conclusions}\label{secConcl}

We have introduced the Complex Hadamard Promise Problem, a family of computational tasks that includes the previously defined Fourier Promise Problem \cite{ACB} and Hadamard Promise Problem \cite{Taddei} as limiting cases, and shown how it can be solved using the quantum switch.  This generalization allows for the definition of promise problems that can be solved in a wider array of physical control and target systems, including continuous variable target systems. Moreover, it uncouples the size $p$ of the Complex Hadamard matrix and the number of quantum gates $N$, requiring only that $p \leq N!$. 

Restricting to the case of finite dimensional target systems, we have shown that these generalized promise problems are restricted to the sub-class of Butson-Hadamard Promise Problems. In that case, the dimension of the target must be compatible with the complexity of the Butson-Hadamard matrix defining the problem. In particular, all Complex Hadamard Promise Problems reduces to the Hadamard Promise Problem when the target system is a qubit. A comparison with related work is summarized in Table I.

Also, we have proved that using a more general class of $(N,p)$-\texttt{switches} to solve the Complex Hadamard Promise Problem provides computational advantages against known fixed-order algorithms in both discrete and continuous variable target systems. To best highlight this advantage, we have introduced  the "query per gate" parameter. The lowest value of this parameter is reached by the quantum switch, which has a fixed $qpg_{switch}=1$, suggesting that the quantum switch could be considered a benchmark for query complexity.

Our work opens the possibility for experimental implementations of promise problems in new platforms, such as CV photonic systems, trapped ions or cavity QED.  Finally, we highlight the necessity of further study on the $(N,p)$-\texttt{switch} as an indefinite causal order instance and its possible application to other computational tasks.

\section*{Acknowledgements}
We thank Daniel Uzcategui and M. Rivera-Tapia for stimulating discussion and reviewing the manuscript. This work was  supported by Fondo Nacional de Desarrollo Cient\'{i}fico y Tecnol\'{o}gico (ANID) (1200266),  ANID – Millennium Science Initiative Program – ICN17\_012, and ANID-Subdirecci\'on de Capital Humano Avanzado/Doctorado Nacional/2021-21211347.

\bibliographystyle{quantum}
\bibliography{refs}

\end{document}